\documentclass{amsart}
\usepackage{amssymb}

\newtheorem{lemma}{Lemma}

\newtheorem{thm}{Theorem}

\def\qed{{\hfill $\square$}}
\def\iH{\mathcal H}
\def\Tr{\mathrm Tr}
\def\<{\langle}
\def\>{\rangle}

\begin{document}
\ \vskip 1cm
\begin{center}{\Large\bf A relation between completely bounded norms and
conjugate channels.} 
\end{center}
\bigskip
\centerline{\large Anna Jen\v cov\'a}
\vskip 3mm
\centerline{\small Mathematical Institute, Slovak Academy of Sciences, Bratislava,Slovakia}
\centerline{\small e-mail:jenca@mat.savba.sk}
\bigskip
\bigskip

\noindent{\bf Abstract.}{\small We show a relation between a quantum channel
$\Phi$ and its conjugate $\Phi^C$, which implies that the $p\to p$ Schatten norm
of the channel is the same as the $1\to p$ completely bounded norm of the
conjugate. This relation is used to give an alternative proof of the
multiplicativity of both norms.}

 \section{Introduction.}
A quantum channel is a completely positive 
trace preserving (CPT) map  $\Phi: M_d\to M_{d'}$, $M_d$ is the set of $d\times
d$ complex matrices. Any channel can 
be  viewed as a  map $L_q(M_d)\to L_p(M_{d'})$,
where $L_q(M_d)$ denotes the space $M_d$ with the Schatten norm
$\|A\|_q=\Tr(|A|^q)^{1/q}$, $1\le q\le\infty$. Let 
$\|\Phi\|_{q\to p}$ be the corresponding norm of $\Phi$, 
$$
\|\Phi\|_{q\to p}=\sup_{A\in M_d}\frac{\|\Phi(A)\|_p}{\|A\|_q}
=\sup_{A\in M_d,A\ge 0}\frac{\|\Phi(A)\|_p}{\|A\|_q},
$$
the second equality was proved in \cite{aud,watrous}. Multiplicativity of this 
 type of
norms is an important conjecture in quantum information theory.

The spaces $L_q(M_d)$ and $L_p(M_{d'})$ can be endowed
with an operator space structure as in  \cite{pisier},
then $\Phi$ is a completely bounded map. 
Multiplicativity of the corresponding completely bounded norms 
$\|\Phi\|_{CB,q\to p}$ for all $1\le p,q\le\infty$ was proved in \cite{DJKR}. 
In particular, this implies 
multiplicativity of $\|\Phi\|_{q\to p}$ for $q\ge p$,  since this is equal to 
$\|\Phi\|_{CB,q\to p}$ for CPT maps. It was shown that the norm $\|\Phi\|_{CB,1\to p}$ is
equal to the quantity
$$
\omega_p(\Phi)=\sup_{\psi\in\mathbb C^d\otimes\mathbb C^d}\frac{\|(\mathcal
I\otimes
\Phi)(|\psi\rangle\langle\psi|)\|_p}{\|\Tr_2(|\psi\rangle\langle\psi|)\|_p}
$$
Multiplicativity of $\omega_p$ then yields
the additivity  for the CB minimal 
conditional entropy, defined as
$$
S_{CB,\min}=\inf_{\psi\in \mathbb C^d\otimes\mathbb C^d}\left(S\left[(\mathcal I
\otimes \Phi)(|\psi\>\<\psi|)\right]-S\left[\Tr_2(|\psi\>\<\psi|)\right]\right)
$$

In the present note, we show that there is a relation between $\omega_p(\Phi)$
and the norm $\|\Phi^C\|_{p\to p}$ of the conjugate map $\Phi^C$. This relation
is then used for an alternative proof of multiplicativity of both quantities, 
avoiding the use of the deep results of the theory of operator spaces and CB
norms, involved in the proofs in \cite{DJKR}.

\section{Representations of CPT maps and conjugate channels.}

Let   $e^d_1,\dots,e^d_d$ be the 
standard basis in $\mathbb C^d$ and let  $\beta_0=\frac 1d\sum_ie^d_i\otimes
e_i^d$  be a maximally entangled vector. Let $\Phi:M_d\to M_{d'}$ be a CPT map. 
Then $\Phi$ is uniquely represented by its Choi-Jamiolkowski matrix 
$X_{\phi}\in M_d\otimes M_{d'}$, defined by
\begin{equation}\label{eq:choi}
X_\Phi=d^2(\mathcal I\otimes
\Phi)(|\beta_0\>\<\beta_0|)= \sum_{i,j}|e^d_i\>\<e^d_j|
\otimes\Phi(|e^d_i\>\<e_j^d|).
\end{equation}
Other representations of $\Phi$ can be obtained from the Stinespring
representation, which in the case of matrices has the form \cite{paulsen}
\begin{equation}\label{eq:stinespring}
\Phi(\rho)= V^\dagger(\rho\otimes I_\kappa)V,\qquad V:\mathbb C^{d'}\to\mathbb
C^d\otimes\iH,\quad \Tr_2VV^\dagger =I_d
\end{equation}
where $\iH$ is an auxiliary Hilbert space,  $\kappa=\dim\iH\le
dd'$. The Lindblad-Stinespring representation
of $\Phi$ is
\begin{equation}\label{eq:lindblad}
\Phi(\rho)=\Tr_2 U(\rho\otimes|\phi\>\<\phi|)U^\dagger
\end{equation}
where $\phi$ is a unit vector in  $\iH$, and $U:\mathbb C^d\otimes\iH\to \mathbb C^{d'}\otimes\iH$ is a partial
isometry. This can be obtained from the Stinespring representation of the dual
map $\hat{\Phi}$. The Kraus representation
\begin{equation}\label{eq:kraus}
\Phi(\rho)=\sum_{k=1}^\kappa F_k \rho F_k^\dagger,\qquad F_k:\mathbb
C^{d}\to\mathbb C^{d'},\quad \sum_k F_k^\dagger
F_k=I_d
\end{equation}
is related to  (\ref{eq:stinespring}) and (\ref{eq:lindblad}) by 
\begin{eqnarray*}
V&=&\sum_{k=1}^{\kappa} F_k^\dagger\otimes |e^\kappa_k\>\\
F_k&=&\Tr_2U(I\otimes|\phi\>\<e^\kappa_k|),\ k=1,\dots,\kappa,
\end{eqnarray*}
where $e^\kappa_1,\dots,e^\kappa_\kappa$ is an orthonormal basis in $\iH$.

Let $\Phi$ be given by (\ref{eq:lindblad}).
The conjugate channel to $\Phi$ is the map $\Phi^C: M_d\to B(\iH)$,
defined as \cite{Holevo,KMNR}
\begin{equation}\label{eq:conjugate}
\Phi^C(\rho)=\Tr_1 U(\rho\otimes|\phi\>\<\phi|)U^\dagger
   =\sum_{j,k}\Tr\left( F_j\rho F_k^\dagger\right )|e^\kappa_j\rangle\langle
   e^\kappa_k|
\end{equation}

The next Lemma shows a relation between the Stinespring representation
(\ref{eq:stinespring}) of $\Phi$ and the Choi-Jamiolkowski
matrix (\ref{eq:choi}) of its conjugate.

\begin{lemma}\label{lemma:relat} Let $\Phi$ be a CPT map, such that
$\Phi(\rho)=V^\dagger(\rho\otimes I_\kappa)V$ is the Stinespring representation. Then
$$X_{\Phi^C}=(VV^\dagger)^T$$
where $B^T$ is the transpose of the matrix $B$, $B^T_{ij}=B_{ji}$.
\end{lemma}

\noindent{\bf Proof:} Let $V=\sum_{k=1}^{\kappa} F_k^\dagger\otimes |
e^\kappa_k\>$,
then using (\ref{eq:conjugate}), we get 
\begin{eqnarray*}
VV^\dagger&=&\sum_{i,j=1}^\kappa F_i^\dagger F_j\otimes |e_i^\kappa\rangle\langle
e_j^\kappa|=
\sum_{i,j=1}^\kappa\sum_{k,l=1}^d \langle e^d_k|F_i^\dagger F_j|e^d_l\rangle\,
|e_k^d\rangle\langle e_l^d|\otimes |e_i^\kappa\rangle\langle e_j^\kappa|\\
&=& \sum_{k,l=1}^d|e_k^d\rangle\langle e_l^d|\otimes
\sum_{i,j=1}^\kappa\Tr \left(F_j|e_l^d\rangle\langle e_k^d|F_i^\dagger\right)
|e_i^\kappa\rangle\langle e_j^\kappa|\\
&=&\sum_{k,l=1}^d|e_k^d\rangle\langle e_l^d|\otimes
\left[\Phi^C(|e_l^d\rangle\langle e_k^d|)\right]^T=X_{\Phi^C}^T
\end{eqnarray*}
\qed

\begin{thm}\label{thm:norms} For a CPT map $\Phi$ and $1\le p\le\infty$,
$$
\|\Phi\|_{p\to p}=\omega_p(\Phi^C)
$$

\end{thm}

\noindent {\bf Proof.} Note first that for any CPT map, we have (\cite{DJKR})
\begin{equation}\label{eq:omega}
\omega_p(\Phi) = \sup_{A\ge 0, \|A\|_{2p}\le 1}\|(A\otimes
I_{d'})X_{\Phi}(A\otimes I_{d'})\|_p,
\end{equation}

Let the Stinespring representation (\ref{eq:stinespring})
of $\Phi$  be $\Phi(\rho)=V^\dagger(\rho\otimes I_\kappa)V$. 
Then by Lemma \ref{lemma:relat},
\begin{eqnarray*}
\|\Phi\|_{p\to p}&=&\sup_{A\ge 0, \|A\|_p\le 1}\|\Phi(A)\|_p=\sup_{B\ge0,
\|B\|_{2p}\le 1}\|V^\dagger(B^2\otimes I_\kappa)V\|_p=\\
&=&\sup_{B\ge0,\|B\|_{2p}\le 1}\|(B\otimes I_\kappa)X_{\Phi^C}^T(B\otimes 
I_\kappa)\|_p=\omega_p(\Phi^C),
\end{eqnarray*}
the last equality follows from the fact that $B^T\ge 0$ if $B\ge 0$ and $\|B^T\|_p=\|B\|_p$.

\qed

\noindent{\bf Remark.} Let $q>p$. Exactly as in the above proof, we get
that
$$
\|\Phi\|_{q\to p}=\sup_{A\ge 0, \|A\|_{2q}\le 1}\|(A\otimes
I_{d'})X_{\Phi^C}(A\otimes I_{d'})\|_p,
$$
The last expression is equal to the $L_r(M_d,L_p(M_{d'}))$ norm 
$\|X_{\Phi^C}\|_{(r,p)}$ for $\frac 1q+\frac1r=\frac1p$, see  Eq. (3.18) in
\cite{DJKR}. This is an operator space type of norm, but not a CB norm, in
general.

\section{Multiplicativity.}

To prove multiplicativity, we  need the following observation
\begin{equation}\label{eq:omegamult}
\omega_p(\Phi\otimes\Tr)=\omega_p(\Tr\otimes\Phi)=\omega_p(\Phi)
\end{equation}
This follows from Lemma \ref{lemma:king}, proved in the Appendix.
We remark that this equality   implies that the supremum in the definition of 
$\omega_p$ can be taken over all 
$M_{d}\otimes M_{d}$, that is,
\begin{equation}\label{eq:omegap}
\omega_p(\Phi)=\sup_{X\in M_{d}\otimes M_{d}}\frac{\|(\mathcal
I\otimes
\Phi)(X)\|_p}{\|\Tr_2(X)\|_p}
\end{equation}
To show this, we first note that the supremum in (\ref{eq:omegap}) may be 
restricted to positive $X$.
Let $X\ge0$ and let $|\psi_{123}\rangle\in \mathbb C^d\otimes\mathbb C^d\otimes \mathbb C^{d^2}$ 
be a purification of $X$, $X=\Tr_3(|\psi_{123}\rangle\langle\psi_{123}|)$.
Then
\begin{eqnarray*}
\frac{\|(\mathcal I\otimes\Phi)(X)\|_p}{\|\Tr_2(X)\|_p}&=&\frac{\|(\mathcal
I_1\otimes
\Phi)(\Tr_3(|\psi_{123}\rangle\langle\psi_{123}|))\|_p}{\|\Tr_{23}(|\psi_{123}\rangle\langle\psi_{123}|)\|_p}=\\ \ \\
&=&\frac{\|(\mathcal
I_1\otimes
\Phi\otimes\Tr)(|\psi_{123}\rangle\langle\psi_{123}|))\|_p}{\|\Tr_{23}(|\psi_{123}\rangle\langle\psi_{123}|)\|_p}
\end{eqnarray*}
Consequently,
\begin{eqnarray*}
\omega_p(\Phi)&\le& \sup_{X\in M_{d}\otimes M_{d}}\frac{\|(\mathcal
I\otimes
\Phi)(X)\|_p}{\|\Tr_2(X)\|_p}\le\\
&\le& \sup_{\psi\in(\mathbb C^d\otimes\mathbb C^{d^2})^{\otimes 2}}
\frac{\|(\mathcal I_{12}\otimes\Phi\otimes\Tr)(|\psi\rangle\langle\psi|)\|_p}{\|\Tr_{34}(|\psi\rangle\langle\psi|)\|_p}=\omega_p(\Phi\otimes\Tr)=
\omega_p(\Phi),
\end{eqnarray*}
hence the assertion.

We now obtain an alternative proof of multiplicativity of $\|\cdot\|_{p\to p}$ and $\omega_p$.

\begin{thm} For CPT maps $\Phi_1:\ M_{d_1}\to M_{d_1'}$ and $\Phi_2:\ M_{d_2}\to
M_{d_2'}$ and for $1\le p\le\infty$,
\begin{eqnarray*}
\|\Phi_1\otimes\Phi_2\|_{p\to p}&=&\|\Phi_1\|_{p\to p}\|\Phi_2\|_{p\to p}\\
\omega_p(\Phi_1\otimes\Phi_2)&=&\omega_p(\Phi_1)\omega_p(\Phi_2)
\end{eqnarray*}
\end{thm}

\noindent{\bf Proof.} We first show that the $p\to p$ norm of a channel $\Phi$ is not changed by tensoring with identity. Indeed,
by Theorem \ref{thm:norms}  and (\ref{eq:omegamult}),
$$
\|\Phi\otimes\mathcal I\|_{p\to p}=\omega_p((\Phi\otimes\mathcal
I)^C)=\omega_p(\Phi^C\otimes\Tr)=\omega_p(\Phi^C)=\|\Phi\|_p
$$
Similarly, $\|\mathcal I\otimes \Phi\|_{p\to p}=\|\Phi\|_{p\to p}$.

Let now $A\in M_{d_1}\otimes M_{d_2}$,  $B =(\mathcal I\otimes \Phi_2)(A)$ 
and compute
$$
\sup_A\frac{\|(\Phi_1\otimes\Phi_2)(A)\|_p}{\|A\|_p}=\sup_A\frac{\|(\Phi_1
\otimes\mathcal I)(B)\|_p}{\|B\|_p}
\frac{\|(\mathcal I\otimes\Phi_2)(A)\|_p}{\|A\|_p}\le \|\Phi_1\|_{p\to p}\|\Phi_2\|_{p\to p}
$$
Since the opposite inequality is easy, we get $\|\Phi_1\otimes \Phi_2\|_{p\to p}=\|\Phi_1\|_{p\to p}\|\Phi_2\|_{p\to p}$, which in turn implies the
 multiplicativity of $\omega_p$.

 \qed
 
\noindent {\bf  Acknowledgements.} This work was done during a visit to 
Tufts University and  thereby partially supported by NSF grant DMS-0314228. 
The author wishes to thank Mary Beth Ruskai and Christopher
King for discussions and valuable comments.
The research was supported by Center of Excellence SAS Physics
of Information I/2/2005 and  Science and Technology Assistance Agency under the
contract  No. APVT-51-032002.

\section*{Appendix.}

The following Lemma is due to C. King.
\begin{lemma}\label{lemma:king}
Let $\Omega :M_n\to M_m$ be a channel with  the covariance property 
$\Omega(U\rho U^\dagger)=U'\Omega(\rho)\bar (U')^\dagger$
where $U'$ is a unitary in $M_m$, for any unitary $U\in M_n$. 
Then for any CPT map $\Phi$, we have
$$
\omega_p(\Omega\otimes\Phi)=\omega_p(\Phi\otimes\Omega)=\omega_p(\Phi)\omega_p(\Omega)
$$
\end{lemma}

\noindent{\bf Proof.} The proof uses the fact that there are $n^2$
unitary operators in $M_n$, such that 
$\sum_{k=0}^{n^2-1}U_kAU_k^\dagger=n(\Tr A) I_n$
for any $n\times n$ matrix $A$, and therefore
$$\sum_k (U_k\otimes
I_d)A_{12}(U_k^\dagger\otimes I_d)
=nI_n\otimes A_2$$ for $A_{12}\in M_n\otimes M_{d}$, $A_2=\Tr_1 A_{12}$.
Let us define
$$
g_p(\rho,\Phi)=\Tr \left[ (\rho^{1/2p}\otimes I_{d'})X_{\Phi}(\rho^{1/2p}\otimes
I_{d'})\right]^p=
\Tr \left[X_\Phi^{1/2}(\rho^{1/p}\otimes I_{d'})X_\Phi^{1/2}\right]^p
$$
so that $\omega_p(\Phi)^p=\sup_{\rho\ge0,\Tr\rho\le1}g_p(\rho,\Phi)$. Then by \cite{epstein},
$\rho\mapsto g_p(\rho,\Phi)$ is concave. It is easy to see that
$g_p(\rho,\Omega)=g_p(U\rho U^\dagger,\Omega)$ for any unitary operator $U$ on
$\mathbb C^n$. It follows that for any $\rho\ge0$, $\Tr\rho=1$
\begin{eqnarray*}
g_p(\rho,\Omega)&=&\frac 1{n^2}\sum_kg_p(U_k\rho U_k^\dagger,\Omega)
\le g_p\left(\frac 1{n^2}\sum_kU_k\rho U_k^\dagger,\Omega\right)\\
&=& g_p(\frac 1n I_n,\Omega)= n^{-1/p}\|X_\Omega\|_p=\omega_p(\Omega)
\end{eqnarray*}
Similarly, we have
\begin{eqnarray*}
g_p(\rho_{12},\Omega\otimes\Phi)&=&
\frac 1{n^2}\sum_kg_p((U_k\otimes I_d)\rho_{12}(U_k^\dagger\otimes I_d),
\Omega\otimes\Phi)\\
&\le& g_p\left(\frac 1{n^2}\sum_k(U_k\otimes I_d)\rho_{12}(U_k^\dagger\otimes
I_d),\Omega\otimes\Phi\right)\\
&=& g_p\left(\frac{1}{n}I_n\otimes \rho_2,\Omega\otimes\Phi\right)=
g_p(\frac 1nI_n,\Omega)g_p(\rho_2,\Phi)
\end{eqnarray*}
The easy inequality $\omega_p(\Omega)\omega_p(\Phi)\le\omega_p(\Omega\otimes\Phi)$ now finishes the proof.
The equality $\omega_p(\Phi\otimes\Omega)=\omega_p(\Phi)\omega_p(\Omega)$ is proved similarly.
\qed

\end{document}